ORIGINAL RESEARCH ARTICLE

# A Fast and Robust Camera-IMU Online Calibration Method For Localization System

Xiaowen Tao, Pengxiang Meng, Bing Zhu*, Jian Zhao

Jilin University, Changchun, China

**ABSTRACT**

Autonomous driving has spurred the development of sensor fusion techniques, which combine data from multiple sensors to improve system performance. In particular, localization system based on sensor fusion , such as Visual Simultaneous Localization and Mapping (VSLAM), is an important component in environment perception, and is the basis of decision-making and motion control for intelligent vehicles. The accuracy of extrinsic calibration parameters between camera and IMU has significant effect on the positioning precision when performing VSLAM system. Currently, existing methods are time-consuming using complex optimization methods and sensitive to noise and outliers due to off-calibration, which can negatively impact system performance. To address these problems, this paper presents a fast and robust camera-IMU online calibration method based space coordinate transformation constraints and SVD (singular Value Decomposition) tricks. First, constraint equations are constructed based on equality of rotation and transformation matrices between camera frames and IMU coordinates at different moments. Secondly, the external parameters of the camera-IMU are solved using quaternion transformation and SVD techniques. Finally, the proposed method is validated using ROS platform, where images from the camera and velocity, acceleration, and angular velocity data from the IMU are recorded in a ROS bag file. The results showed that the proposed method can achieve robust and reliable camera-IMU online calibration parameters results with less tune consuming and less uncertainty.

**KEYWORDS:** Camera-IMU Calibration; Autonomous Vehicles; Sensor Fusion; SLAM; Localization System

## Introduction

The rapid development of artificial intelligence has sparked a growing interest in autonomous driving. With breakthroughs in machine learning and advanced sensor technologies, autonomous vehicles have become a hot topic in recent years. This revolutionary technology has the potential to transform transportation by enhancing safety, improving efficiency, and reducing traffic congestion [1].

Autonomous driving encompasses a complex system comprising perception, localization, decision-making, planning, and control. By integrating these components, self-driving vehicles are poised to revolutionize transportation. Particularly, the localization system utilizes various sensors to accrately comprehend the surrounding environment and achieve high-precision positioning, which is then transmitted to the planning and decision-making systems. The vehicle's precise localization information not only serves as a prerequisite for global and local decision-making and trajectory planning but also influence the calculation of parameters in the control execution system. Thus, the demand for high-precision vehicle localization is a crucial foundation for correct decision-making, planning, and control execution in intelligent vehicles. Additionally, as intelligent vehicles play a vital role in the Intelligent Transportation System (ITS), their localization technology holds significant importance in constructing the ITS. For example, improved localization precision in intelligent vehicles enhances the accuracy of real-time traffic information in the ITS, enabling more rational selection of travel routes and dynamic path guidance. Therefore, the breakthrough in high-precision localization technology for intelligent vehicles holds great significance for the automotive industry's substantial development, improving the quality of life, reducing environmental pollution, and enhancing the economy [2].

Currently, common positioning methods in intelligent vehicles include Global Positioning System (GPS), Ultra Wide Band (UWB) positioning, LiDAR positioning, Radio Frequency Identification (RFID) positioning, visual positioning, Real-Time Kinematic (RTK) positioning, Inertial Measurement Unit (IMU) positioning, Dead Reckoning (DR) map matching, etc. Although there are numerous existing positioning methods, each method inevitably has its limitations and lacks generalization capabilities [1-3]. For instance, absolute positioning methods offer higher accuracy compared to relative positioning methods, but they are more expensive and sensitive to obstacles and interference. Relative positioning methods rely on past sensor information to estimate the vehicle's pose at the next moment, leading to inevitable error accumulation, making them unsuitable for long-term positioning requirements. Therefore, to enhance the accuracy, stability, and generalization capabilities of positioning systems, it is necessary to leverage multi-sensor fusion technology to exploit the advantages of different sensors. Currently, multi-sensor fusion-based positioning technology is a research hotspot in intelligent vehicle positioning. Common fusion positioning techniques include GPS/IMU, vision/IMU, GPS/vision, GPS/DR, LiDAR/IMU, LiDAR/GPS, LiDAR/GPS/RTK, etc.

IMU positioning refers to obtaining velocity, position, and rotation information using built-in three-axis gyroscopes and three-axis accelerometers [3]. It relies on internal sensors and does not depend on external environments, providing high positioning accuracy in a short time. However, IMU sensors, including accelerometers and gyroscopes, have biases, and over time, measurement errors accumulate, making them unsuitable for long-term positioning requirements. Visual positioning does not suffer from drift and can directly measure rotation and translation, which is commonly used to calibrate the errors in IMU measurements using low-cost camera sensors. Additionally, IMU can estimate absolute scale and respond to rapid motions and rotations, compensating for the drawbacks of visual positioning, such as the inability to measure scale and the loss of measurement information during fast movements [4]. Therefore, Camera-IMU localization system is currently a mainstream sensor-fusion positioning method due to its advantages of low cost, high efficiency, high accuracy, convenience, and speed.

In Camera-IMU localization system, camera and IMU calibration stands out as a critical process for accurate data integration. By aligning visual information from cameras with inertial measurements from IMUs, this calibration ensures precise and synchronized perception for robust decision-making in autonomous systems. Calibration plays a crucial role in sensor fusion, encompassing camera intrinsic parameters, IMU intrinsic parameters, and camera-to-IMU extrinsic calibration. First, camera intrinsic parameters describe the projection relationship between the three-dimensional real world and the camera coordinate system. They play a critical role in ensuring the accuracy of the camera's data source. These parameters include focal length, principal point, and distortion coefficients, which directly impact the quality and accuracy of the captured images. Secondly, IMU intrinsic parameter calibration mainly involves parameters related to deterministic errors and random errors. Deterministic errors refer to systematic biases or scale factors that affect the IMU measurements consistently. These errors can arise from factors such as sensor misalignments, temperature variations, and nonlinearity in sensor response. Calibration helps estimate and compensate for these deterministic errors, improving the accuracy and reliability of the IMU measurements. Random errors, on the other hand, are typically caused by noise and uncertainties in the IMU sensors. These errors are characterized by their statistical properties, such as zero mean and Gaussian distribution. IMU calibration aims to characterize and minimize these random errors to enhance the accuracy and precision of the IMU measurements. Camera-to-IMU extrinsic parameters establish the transformation between the camera and IMU measurements, aligning them to a common coordinate frame. These parameters are crucial for Visual Simultaneous Localization and Mapping (VSLAM) systems. Even with a deviation of just 1° to 2° between the calibrated camera sensor coordinate system and the IMU coordinate system, the localization accuracy of VSLAM system can still degrade significantly [4-5]. Therefore, the calibration of cameras and IMUs is a critical step in sensor fusion-based positioning systems, ensuring the accuracy and effectiveness of the fusion process and ultimately improving the performance and reliability of the overall positioning system.

Curretnly, there are several research to make sure the accurate calibration between camera and IMU. Regarding intrinsic calibration, Zhang et al. [6] introduced a widely used camera calibration technique known as Zhang's method. This method employs a calibration pattern with known geometry and utilizes multiple images to estimate the camera's intrinsic parameters. Further advancements in camera calibration techniques include self-calibration methods [7] and techniques based on pattern recognition and optimization algorithms [8]. One common approach for extrinsic calibration is the use of a calibration target and a series of controlled movements. Zhang et al. [9] proposed a method using a planar calibration target to estimate the transformation matrix between the camera and IMU. They employed bundle adjustment to optimize the calibration results. Similarly, Brink et al. [10] utilized a multi-view calibration target to estimate the relative pose between the camera and IMU. They employed a Kalman filter-based optimization approach for accurate extrinsic calibration.

Despite these advancements, Camera-IMU calibration techniques still face limitations. Sünderhauf et al. [11] highlighted challenges such as the sensitivity to initialization, the need for precise sensor synchronization, and the effects of environmental factors on calibration accuracy. These factors can lead to errors in the calibration process and subsequently affect the accuracy of the sensor fusion system. In addition, existing methods are time-consuming using complex optimization methods and sensitive to noise and outliers due to off-calibration, which can negatively impact system performance.

In this paper, we present a fast and robust camera-IMU online calibration method based space coordinate transformation constraints and SVD (singular Value Decomposition) tricks. First, constraint equations are constructed based on equality of rotation and transformation matrices between camera frames and IMU coordinates at different moments. Secondly, the external parameters of the camera-IMU are solved using quaternion transformation and SVD techniques. The remainder of this paper is organized as follows: Section 1 introduces the calibration for camera's parameters. The details of the IMU's parameters calibration are described in Section 2. The innovative Caemera-IMU extrinsic parameter calibration is given in Section 3. Experimental results can be found in Section 4. Finally, the conclusions are presented in Section 5.

# 1. Calibration of Camera Intrinsic Parameters Based on Pinhole Model

## 1.1 Pinhole Model

Before extracting image features for positioning, it is necessary to model the camera sensor to obtain accurate and stable input image frames. An image frame is a two-dimensional image plane composed of pixels with pixel coordinates and brightness information, while corresponding landmarks exist in three-dimensional space. Therefore, a camera model is used to describe the one-to-one mapping between three-dimensional landmarks and the two-dimensional image plane. There are various camera models, including the pinhole camera model [12], stereo camera model [13], RGB-D camera model [5], among others. Among these models, the most commonly used and simple yet effective one is the pinhole camera model, which sufficiently meets the requirements of visual-inertial odometry systems for camera sensors. Thus, this paper employs the pinhole camera model to describe the entire mapping process.

The imaging principle of a camera sensor is depicted in Figure 1-2.

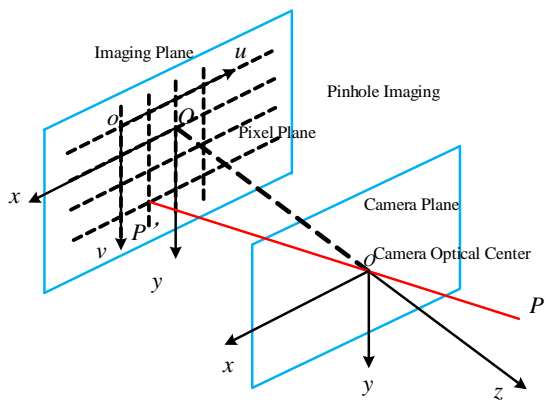 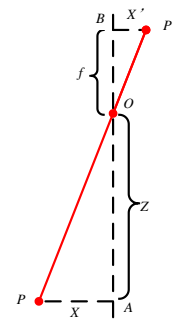

Fig. 1. Schematic diagram of the pinhole camera model.　　　　Fig. 2. Schematic diagram of similar triangles.

The pinhole camera model depicted in Figure 1 establishes a camera coordinate system, denoted as *Oxyz*, with the camera's optical center $O$ as the origin. The *x*-axis is aligned with the right side of the camera sensor, the *y*-axis is aligned with the vertical direction of the camera sensor, and the *z*-axis is aligned with the front direction of the camera sensor. Similarly, let's establish a camera image plane coordinate system denoted as *O'x'y'*. Assuming a point $P = [X, Y, Z]^{\mathrm{T}}$ in the three-dimensional world projects through the pinhole O onto the image plane as point $P' = [X', Y', Z']^{\mathrm{T}}$, and considering the distance between the camera plane and the physical image plane as the focal length f. According to the principle of similar triangles in the pinhole camera model depicted in Figure 2, we can derive the following equation:

$$\frac{Z}{f} = -\frac{X}{X'} = -\frac{Y}{Y'} \tag{1}$$

The negative sign in Equation (1) indicates that the resulting image is inverted. However, in reality, the image captured by a camera sensor is not inverted. To better represent the pinhole camera model and align with the actual scenario, this paper symmetrically places the camera image plane in front of the camera sensor, as illustrated in Figure 3.

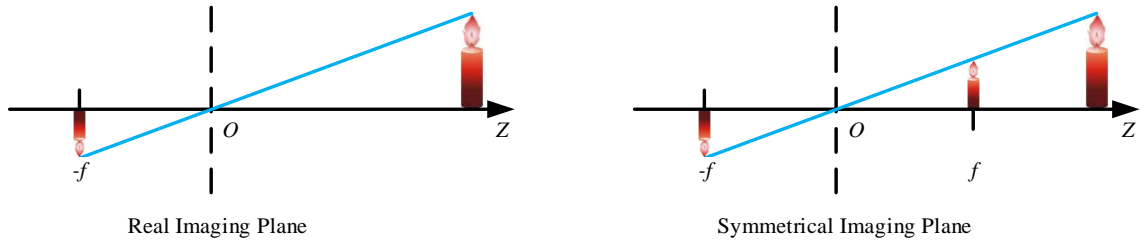

Real Imaging Plane          Symmetrical Imaging Plane

Fig. 3. Schematic diagram of both the actual image plane and the symmetric image plane in the pinhole camera model.

By removing the symbols in Equation (1) and rearranging the terms, we can obtain the Equation (2):

$$\begin{cases} X' = f\dfrac{X}{Z} \\ Y' = f\dfrac{Y}{Z} \end{cases} \tag{2}$$

Equation (2) describes the three-dimensional spatial relationship between the 3D landmark point $P$ and its image point $P'$. However, the information obtained from the camera sensor is in the form of individual pixels. Therefore, it is necessary to perform translation and scaling on the image plane coordinate system $O'x'y'$ to obtain the pixel coordinate system $ouv$, as shown in Figure 1. Assuming that the pixel coordinate system is scaled by $\alpha$ and $\beta$ along the u and v axes respectively, and translated by $c_x$ and $c_y$, we can establish the relationship between the image point $P'$ and its pixel coordinates $[u, v]^T$ as follows:

$$\begin{aligned} u &= \alpha X' + c_x \\ v &= \beta Y' + c_y \end{aligned} \tag{3}$$

Let $f_x = \alpha f$ and $f_y = \beta f$, we can rearrange the equation to obtain Equation (4):

$$\begin{cases} u = f_x \dfrac{X}{Z} + c_x \\ v = f_y \dfrac{Y}{Z} + c_y \end{cases} \tag{4}$$

Expressing Equation (4) in matrix form and simplifying, we have:

$$Z\begin{bmatrix} u \\ v \\ 1 \end{bmatrix} = \begin{bmatrix} f_x & 0 & c_x \\ 0 & f_y & c_y \\ 0 & 0 & 1 \end{bmatrix} \begin{bmatrix} X \\ Y \\ Z \end{bmatrix} \triangleq \boldsymbol{KP} \tag{5}$$

Where the matrix composed of the intermediate quantities is the camera's intrinsic parameter matrix $K$. Based on coordinate transformation knowledge, the coordinates in the camera coordinate system and the coordinates in the world coordinate system $P_w$ have the following relationship:

$$P = RP_w + t \tag{6}$$

Where $R$ and $t$ are the rotation matrix and translation matrix that transform coordinates from the world coordinate system to the camera coordinate system. The corresponding coordinates $P_{uv}$ in the pixel coordinate system can be related to the world coordinate system, camera coordinate system, and pixel coordinate system as follows:

$$ZP_{uv} = Z\begin{bmatrix} u \\ v \\ 1 \end{bmatrix} = K(RP_w + t) = KTP_w \tag{7}$$

Where $T = [R, \ t]$ is the transformation matrix. In Equation (7), it specifically represents the transformation matrix from the world coordinate system to the pixel coordinate system. To facilitate the derivation of the camera sensor's intrinsic parameters, it is common to normalize Equation (7) by setting Z=1 and perform calculations based on the projection on the normalized plane, as illustrated in Figure 4.

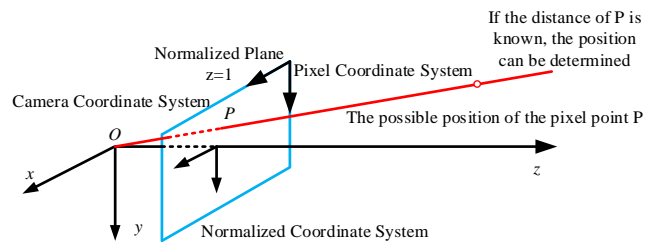

Fig. 4. Schematic diagram of the normalized plane.

This paper utilizes two constrains to solve the intrinsic parameter. First, when rotating cameras, the two rotation vectors along the camera center are orthogonal to each other. Secondly, the magnitude of the rotation vector is 1. According that, we can obtain the parameter's value in the intrinsic parameter matrix, as shown below.

$$\begin{cases} f_x = \sqrt{\lambda/B_{11}} \\ f_y = \sqrt{\lambda B_{11}/(B_{11}B_{22} - B_{12}^2)} \\ c_x = \gamma c_y/f_y - B_{13}f_x^2/\lambda \\ c_y = (B_{12}B_{13} - B_{11}B_{23})/(B_{11}B_{22} - B_{12}^2) \\ \gamma = -B_{12}f_x^2 f_y/\lambda \\ \lambda = B_{33} - [B_{13}^2 + c_y(B_{12}B_{13} - B_{11}B_{23})]/B_{11} \end{cases} \tag{8}$$

The $B$ matrix is a matrix defined by us, and it is defined as follows:

$$B = (K^{-1})^T K^{-1} = \begin{bmatrix} B_{11} & B_{12} & B_{13} \\ B_{21} & B_{22} & B_{23} \\ B_{31} & B_{32} & B_{33} \end{bmatrix} \tag{9}$$

## 1.2 Distortion Model

Additionally, due to the lens on the camera sensor, the projection from the three-dimensional world to the two-dimensional image plane introduces distortion [14] caused by the optical properties of the lens. Therefore, this paper also incorporates a distortion model to describe the projection process from the three-dimensional world to the two-dimensional image plane.

Typically, a lens is placed in front of the camera, resulting in radial distortion, as shown in Figure 5. Moreover, during the hardware assembly process, it is inevitable that the lens and the imaging plane are not perfectly aligned, leading to tangential distortion, as depicted in Figure 6.

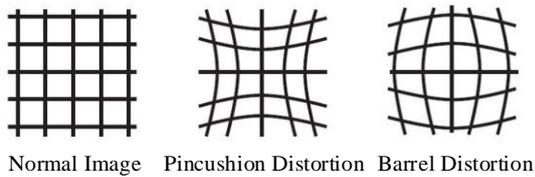
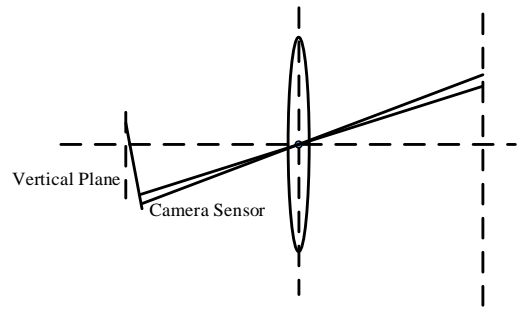

Fig. 5. Schematic diagram of radial distortion.      Fig. 6. Schematic diagram of tangential distortion source.

Therefore, in order to obtain better imaging results, it is necessary to perform distortion correction on the camera sensor. Assuming that tangential distortion and radial distortion follow a polynomial relationship, for any point $p = [x, y]^T$ on the normalized plane, the corrected normalized coordinates $[x_{\text{dis\_r}}, y_{\text{dis\_r}}]^T$ after radial distortion correction are related as follows:

$$x_{\text{dis\_r}} = x(1 + k_1 r^2 + k_2 r^4) \tag{10}$$

$$y_{\text{dis\_r}} = y(1 + k_1 r^2 + k_2 r^4) \tag{11}$$

Where $k_1$ and $k_2$ are radial distortion coefficients, and $r$ is the distance between point $p$ and the coordinate origin in polar coordinates. Similarly, for tangential distortion, the corrected normalized coordinates $[x_{\text{dis\_t}}, y_{\text{dis\_t}}]^T$ of point p can be related as follows:

$$x_{\text{dis\_t}} = x + 2p_1 xy + p_2(r^2 + 2x^2) \tag{12}$$

$$y_{\text{dis\_t}} = y + p_1(r^2 + 2y^2) + 2p_2 xy \tag{13}$$

Where $p_1$ and $p_2$ are tangential distortion coefficients, and $r$ is also the distance of point $p$ from the origin in polar coordinates. Therefore, by combining the operations of radial and tangential distortion correction on points in the normalized plane, we obtain the distortion correction parameter matrix $U_d = \begin{bmatrix} k_1 & k_2 \\ p_1 & p_2 \end{bmatrix}$. Consequently, we can obtain the normalized coordinates of the undistorted points $[x_{\text{distorted}}, y_{\text{distorted}}]^T$ as follows:

$$x_{\text{distorted}} = x(1 + k_1 r^2 + k_2 r^4) + 2p_1 xy + p_2(r^2 + 2x^2) \tag{14}$$

$$y_{\text{distorted}} = y(1 + k_1 r^2 + k_2 r^4) + p_1(r^2 + 2y^2) + 2p_2 xy \tag{15}$$

For the distortion correction coefficients, rearranging them into the distortion correction parameter matrix, and substituting Equation (14)-(15) into Equation (4), we can derive the expression for projecting the undistorted points onto the pixel plane using the intrinsic parameter matrix. It is given as follows:

$$\begin{cases} u = f_x x_{\text{distorted}} + c_x \\ v = f_y y_{\text{distorted}} + c_y \end{cases} \tag{16}$$

By utilizing the pinhole camera model and distortion model, the three-dimensional landmarks can be projected onto the camera's two-dimensional image plane, enabling the estimation of the camera's intrinsic parameters.

## 2. IMU Intrinsic Parameters Calibration

### 2.1 IMU Measurement Model

The measurement errors of an IMU mainly consist of Gaussian white noise and biases [1,15]. Therefore, this paper considers the random walk of biases and noise in modeling the IMU's measured acceleration $\hat{a}_t^b$ and angular velocity $\hat{\omega}_t^b$:

$$\hat{a}_t^b = a_t^b + R_{bw} g^w + b_{a_t} + n_a \tag{17}$$

$$\hat{\omega}_t^b = \omega_t^b + b_{\omega_t} + n_\omega \tag{18}$$

Where $R_{bw}$ is the rotation matrix from the world coordinate system to the IMU coordinate system; $g^w = [0, 0, g]^T$ is the gravity vector in the world coordinate system; $b_{a_t}$ is the acceleration bias; $n_a$ is the noise of acceleration; $b_{\omega_t}$ is the bias of angular velocity; $n_\omega$ is the noise of angular velocity. This paper assumes that the acceleration noise $n_a$ and the angular velocity noise $n_\omega$ obey the Gaussian distribution, that is, $n_a \sim \mathcal{N}(0, \sigma_a^2)$, $n_\omega \sim \mathcal{N}(0, \sigma_\omega^2)$. Assume that the acceleration deviation $b_{a_t}$ and the deviation $b_{\omega_t}$ from the gyroscope are modeled as a bounded random walk and a random walk respectively, and its first derivative is as follows:

$$\dot{b}_{a_t} = -\frac{1}{\tau} b_{a_t} + n_{b_a} \tag{19}$$

$$\dot{b}_{\omega_t} = n_{b_\omega} \tag{20}$$

Where $\tau$ is the time constant; $n_{b_a}$ and $n_{b_\omega}$ are the noise of the acceleration bias and the noise of the gyroscope bias respectively, which obey the Gaussian distribution, that is, $n_{b_a} \sim \mathcal{N}(0, \sigma_{b_a}^2)$, $n_{b_\omega} \sim \mathcal{N}(0, \sigma_{b_\omega}^2)$.

### 2.2 IMU Errors

IMU intrinsic parameters calibration mainly involves parameters related to random errors and deterministic errors. The parameters related to random errors are nois and random walk. In the actual process of imu sampling, discretization is required, so this paper deduces the relationship between the continuation and discretization of random errors [3,16].

Integrating white Gaussian noise over a fixed time interval $\Delta t$, we can write:

$$n_d(k) = n(t + \Delta t) \approx \frac{1}{\Delta t} \int_t^{t+\Delta t} n(\tau)\, dt \tag{21}$$

Where the covariance is:

$$E(n_d^2(k)) = E\left(\frac{1}{\Delta t^2} \int_{t_0}^{t_0+\Delta t} \int_{t_0}^{t_0+\Delta t} n(\tau) n(t)\, d\tau dt \right) = \frac{\sigma^2}{\Delta t} \tag{22}$$

It can be seen from Equation (22) that the variance of the discrete Gaussian white noise can be obtained by simply dividing the continuous Gaussian white noise variance by a time interval. Similarly, for the integral of a bias random walk over a fixed time interval $\Delta t$, we can write:

$$b(t_0 + \Delta t) = b(t_0) + \int_{t_0}^{t_0+\Delta t} n_{b_\omega}(t)\, dt \tag{23}$$

Where the covariance is:

$$E\big(b^2(t_0 + \Delta t)\big) = E\left(\left[b(t_0) + \int_{t_0}^{t_0+\Delta t} n_{b_\omega}(t)\, dt\right]\left[b(t_0) + \int_{t_0}^{t_0+\Delta t} n_{b_\omega}(\tau)\, d\tau\right]\right) = \sigma_b^2 \Delta t \tag{24}$$

Similarly, it can be seen from Equation (22) that the variance of the discrete Gaussian white noise can be obtained by simply timing the continuous Gaussian white noise variance by a time interval.

The parameters related to deterministic error are mainly bias, scale and misalignment coefficient. In this paper, the parameters to be optimized are obtained by constructing the loss functions of accelerator and gyroscope respectively.

$$L(\boldsymbol{\theta}_{acc}) = \sum_{k=1}^{M} \left(\|\boldsymbol{g}\|^2 - \left\|h(\boldsymbol{a}_k^b, \boldsymbol{\theta}_{acc})\right\|^2\right)^2 \tag{25}$$

$$L(\boldsymbol{\theta}_{gyro}) = \sum_{k=2}^{M} \left\|\boldsymbol{u}_{a,k} - \boldsymbol{u}_{g,k}\right\|^2 \tag{26}$$

Where $\boldsymbol{g}$ is the gravity vector; $h(\boldsymbol{a}_k^b, \boldsymbol{\theta}_{acc})$ describes the relationship between the acceleration of IMU $\boldsymbol{a}_k^b$ and the real acceleration in the world given the estimated accelerators' parameters $\boldsymbol{\theta}_{acc}$; $\boldsymbol{\theta}_{gyro}$ is the esitmated gyroscopes' parameters; $M$ is the accounted interval times for computing; $\boldsymbol{u}_{a,k}$ is the initial gravity given by calibrated accelerator; $\boldsymbol{u}_{g,k}$ is the calculated gravity, as shown below.

$$\boldsymbol{u}_{g,k} = \psi[\boldsymbol{\omega}_k^b, \boldsymbol{u}_{a,k-1}] \tag{27}$$

## 3. Camera-IMU Online Calibration

### 3.1 IMU Pre-intergration Model

Since the sampling frequency of the IMU is much higher than the sampling frequency of the camera sensor, this paper integrates the IMU measurements between two consecutive image frames to calculate the position, velocity and rotation, so as Insight - Automatic Control(2023.05)to achieve the alignment of the IMU and camera sampling frequency. The schematic diagram of IMU pre-intergration model can be found in Figure 7.

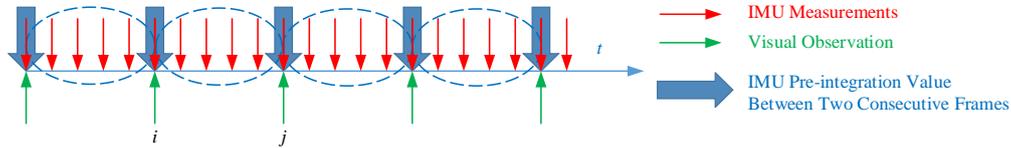

Fig. 7. Schematic diagram of IMU pre-integration model

According to the IMU measurement model equations in Section 2.1, the position, speed, and rotation information in the world coordinate system can be calculated, as follows:

$$\boldsymbol{p}_{wb_j} = \boldsymbol{p}_{wb_i} + \boldsymbol{v}_i^w \Delta t_i + \iint_{t \in [i,j]} (\boldsymbol{R}_{wb_t}(\hat{\boldsymbol{a}}_t^b - \boldsymbol{b}_{a_t} - \boldsymbol{n}_a) - \boldsymbol{g}^w) dt^2 \tag{28}$$

$$\boldsymbol{v}_{b_j}^w = \boldsymbol{v}_{b_i}^w + \int_{t \in [i,j]} (\boldsymbol{R}_{wb_t}(\hat{\boldsymbol{a}}_t^b - \boldsymbol{b}_{a_t} - \boldsymbol{n}_a) - \boldsymbol{g}^w) dt \tag{29}$$

$$\boldsymbol{q}_{wb_j} = \int_{t \in [i,j]} \boldsymbol{q}_{wb_t} \otimes \begin{bmatrix} 0 \\ \frac{1}{2}\boldsymbol{\omega}^{b_t} \end{bmatrix} dt \tag{30}$$

Where $\boldsymbol{p}_{wb_j}$ and $\boldsymbol{p}_{wb_i}$ are the translation from the IMU coordinate system to the world coordinate system at time $j$ and time $i$ respectively; $\boldsymbol{v}_{b_j}^w$ and $\boldsymbol{v}_{b_i}^w$ are the IMU's speed information in the world coordinate system at time $j$ and time $i$ respectively; $\boldsymbol{q}_{wb_j}$ and $\boldsymbol{q}_{wb_i}$ are the rotation quaternions from the IMU coordinate system to the world coordinate system at time $j$ and time $i$ respectively; $\Delta t_i$ is the time interval between the $i$ and $j$ image frames ; $\otimes$ is the multiplication operation of quaternion.

According to the rotation relationship of the coordinate system, this paper rotates the equation (28)-(30) from the world coordinate system to the IMU coordinate system of the $i$-th frame at the same time, so the IMU pre-integration formula at continuous moments can be obtained:

$$\boldsymbol{R}_{b_i w} \boldsymbol{p}_{wb_j} = \boldsymbol{R}_{b_i w} \left( \boldsymbol{p}_{wb_i} + \boldsymbol{v}_i^w \Delta t_i - \frac{1}{2} \boldsymbol{g}^w \Delta t_i^2 \right) + \boldsymbol{\alpha}_{b_i b_j} \tag{31}$$

$$\boldsymbol{R}_{b_i w} \boldsymbol{v}_{b_j}^w = \boldsymbol{R}_{b_i w} (\boldsymbol{v}_{b_i}^w - \boldsymbol{g}^w \Delta t_i) + \boldsymbol{\beta}_{b_i b_j} \tag{32}$$

$$\boldsymbol{q}_{b_i w} \otimes \boldsymbol{q}_{wb_j} = \boldsymbol{\gamma}_{b_i b_j} \tag{33}$$

The symbols in the above equations are represented as follows:

$$\boldsymbol{\alpha}_{b_i b_j} = \iint_{t \in [i,j]} \boldsymbol{R}_{b_i b_t} (\hat{\boldsymbol{a}}_t^b - \boldsymbol{b}_{a_t} - \boldsymbol{n}_a) dt^2 \tag{34}$$

$$\boldsymbol{\beta}_{b_i b_j} = \int_{t \in [i,j]} \boldsymbol{R}_{b_i b_t} (\hat{\boldsymbol{a}}_t^b - \boldsymbol{b}_{a_t} - \boldsymbol{n}_a) dt \tag{35}$$

$$\boldsymbol{\gamma}_{b_i b_j} = \int_{t \in [i,j]} \frac{1}{2} \boldsymbol{\Omega}(\hat{\boldsymbol{\omega}}_t^b - \boldsymbol{b}_{\omega_t} - \boldsymbol{n}_\omega) \boldsymbol{\gamma}_{b_i b_t} dt \tag{36}$$

For $\boldsymbol{\Omega}(\boldsymbol{\omega})$ in the Equation (36), the definition in this paper is as follows:

$$\boldsymbol{\Omega}(\boldsymbol{\omega}) = \begin{bmatrix} -[\boldsymbol{\omega}]^\wedge & \boldsymbol{\omega} \\ -\boldsymbol{\omega}^T & 0 \end{bmatrix} \tag{37}$$

$$[\boldsymbol{\omega}]^\wedge = \begin{bmatrix} 0 & -\omega_z & \omega_y \\ \omega_z & 0 & -\omega_x \\ -\omega_y & \omega_x & 0 \end{bmatrix} \tag{38}$$

Where $\omega_x$, $\omega_y$, and $\omega_z$ are the components of $\boldsymbol{\omega}$ on the corresponding coordinate axes, respectively.

## 3.2 Camera-IMU Extrinsic Parameters Solving

The extrinsic parameters of the camera and IMU refer to the transformation matrix between the camera coordinate system and the IMU coordinate system. The external parameter calibration of the camera and the IMU is very important for the VSLAM system. Even if the final camera coordinate system and the IMU coordinate system are calibrated with only a 1-2° deviation, the positioning accuracy of the VSLAM system will also decrease to be very bad. Therefore, it is necessary to calibrate the external parameters of the camera and IMU online, and the coordinate relationship between the camera and the IMU is shown in Figure 8.

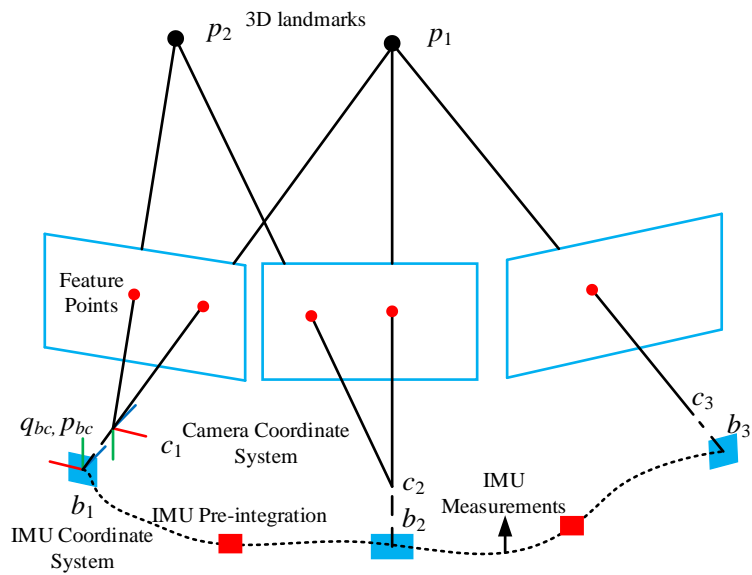

Fig. 8. Schematic diagram of coordinate relationship between the camera and the IMU.

According to the IMU pre-integration model in Section 3.1, the IMU rotation matrix $\boldsymbol{R}_{b_k b_{k+1}}$ from time *k*+1 to time *k* is obtained, and the camera frame at any moment satisfies the following equation:

$$\boldsymbol{R}_{b_k b_{k+1}} \boldsymbol{R}_{bc} = \boldsymbol{R}_{bc} \boldsymbol{R}_{c_k c_{k+1}} \tag{39}$$

Where $\boldsymbol{R}_{bc}$ is the rotation matrix transformed from the camera coordinate system to the IMU coordinate system; $\boldsymbol{R}_{c_k c_{k+1}}$ is the camera rotation matrix from time *k*+1 to time *k*. Write Equation (39) in the form of quaternion:

$$q_{b_k b_{k+1}} \otimes q_{bc} = q_{bc} \otimes q_{c_k c_{k+1}} \tag{40}$$

According to the operational properties of quaternions, the Equation (40) is transformed into the form of quaternion left multiplication matrix and right multiplication matrix, as shown below.

$$\left([q_{b_k b_{k+1}}]_L - [q_{c_k c_{k+1}}]_R\right) q_{bc} = Q_{k+1}^k q_{bc} = 0 \tag{41}$$

Where $[q_{b_k b_{k+1}}]_L$ and $[q_{c_k c_{k+1}}]_R$ are the left multiplication matrix and right multiplication matrix of quaternion respectively. For *N* pairs of measurements, the following system of equation can be constructed:

$$\begin{bmatrix} w_1^0 Q_1^0 \\ w_2^1 Q_2^1 \\ \vdots \\ w_N^{N-1} Q_N^{N-1} \end{bmatrix} q_{bc} = 0 \tag{42}$$

Where $w_{k+1}^k$ is the Huber robust kernel function, which is used to represent the weight of removing outliers, expressed as follows:

$$w_{k+1}^k = \begin{cases} 1, & r_{k+1}^k < r_{thr} \\ \dfrac{r_{thr}}{r_{k+1}^k}, & \text{others} \end{cases} \tag{43}$$

Where $r_{thr}$ is the angle residual threshold of the rotation matrix, which is set to $5°$ in this paper; $r_{k+1}^k$ is the angle residual of the rotation matrix, defined as follows:

$$r_{k+1}^k = arccos\left(\frac{tr(\hat{R}_{bc}^{-1} R_{b_k b_{k+1}}^{-1} \hat{R}_{bc} R_{c_k c_{k+1}}) - 1}{2}\right) \tag{44}$$

So far, the extrinsic parameter $q_{bc}$ of the camera and IMU can be obtained by SVD decomposition of Equation (44).

## 4. Experiments

In this paper, Huawei notebook computer MateBook D is used as the carrier, and the computer configuration is shown in Table 1. The Robot Operating System (ROS)-Kinetic[3] version in the Ubuntu 16.04 system is used as the experimental operating environment, and the proposed Camera-IMU online calibration method is realized through C++ programming[1].

Table 1 Test computer equipment parameters

| Configuration Type | Configuration Information |
|---|---|
| CPU | i5-1135G7 Processor |
| CPU Graphics Card | Intel Integrated Graphics |
| Processor Base Frequency | 2.4GHz |
| Running Memory | 16GB |

The MYNT EYE Camera D1010-50 integrates the camera and IMU, as shown in Figure 9. This paper only uses the left-eye camera and IMU sensor in the MYNT EYE camera. The definition of the camera sensor and IMU coordinate system is shown in Figure 10.

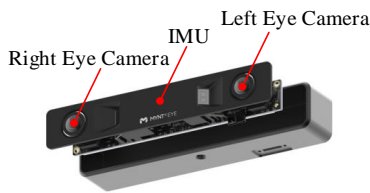 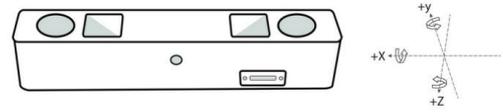

Fig. 9. Schematic diagram of the pinhole camera model.   Fig. 10. Schematic diagram of similar triangles.

The coordinate systems of the camera and IMU is defined as follows:

1) Camera: Take the right-right direction of the camera as the x-axis, take the direction directly behind the camera as the y-axis, and take the direction directly above the camera as the z-axis, conforming to the right-handed coordinate system.
2) IMU: The x-axis is the direction directly to the right of the camera, the y-axis is the direction directly in front of the camera, and the z-axis is the direction directly below the camera, conforming to the right-handed coordinate system.

The performance parameter information of MYNT EYE Camera D1010-50 is shown in Table 2.

Table 2 Performance Parameters of MYNT EYE Camera D1010-50

| Configuration Type | Configuration Information |
|---|---|
| Camera Frame Rate | 10Hz |
| Resolution | 1280×720 |
| Focal Length | 3.9mm |
| IMU Frequency | 200Hz |
| Output Data Format | YUYV/MJPG |
| Output Accuracy | ⩽2.5% |

## 4.1 Experiment of Camera Intrinsic Parameter Calibration

In this paper, Aprilgrid calibration boards are used to calibrate the internal parameters of the camera. The number of Apriltags used in this paper is 6×6, where tagSize is 0.021m and tagSpacing is 0.0063m. When calibrating, keep the camera still, move and rotate the calibration board continuously and ensure that the camera captures photos of the calibration board in various postures, as shown in Figure 11.

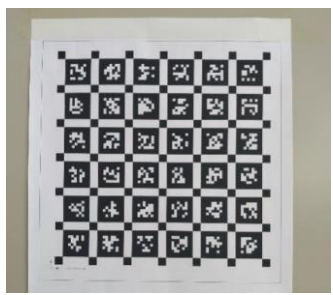

Fig. 11 Aprilgrid calibration boards.

The internal parameter matrix $K$ and distortion correction parameter matrix $U_d$ of the Xiaomi camera are calculated by the method proposed in this paper, as follows:

$$K = \begin{bmatrix} f_x & 0 & c_x \\ 0 & f_y & c_y \\ 0 & 0 & 1 \end{bmatrix} = \begin{bmatrix} 1091.635505127837 & 0 & 615.7646844724167 \\ 0 & 1094.097509334247 & 336.08607722962415 \\ 0 & 0 & 1 \end{bmatrix} \quad (45)$$

$$U_d = \begin{bmatrix} k_1 & k_2 \\ p_1 & p_2 \end{bmatrix} = \begin{bmatrix} 0.0158121998824731 & -0.04906947859369595 \\ -0.007932332725861788 & -0.0036593828274275953 \end{bmatrix} \quad (46)$$

After the camera is calibrated, it is necessary to verify the obtained parameter matrix, and its polar angle and azimuth angle errors are shown in Figure 12.

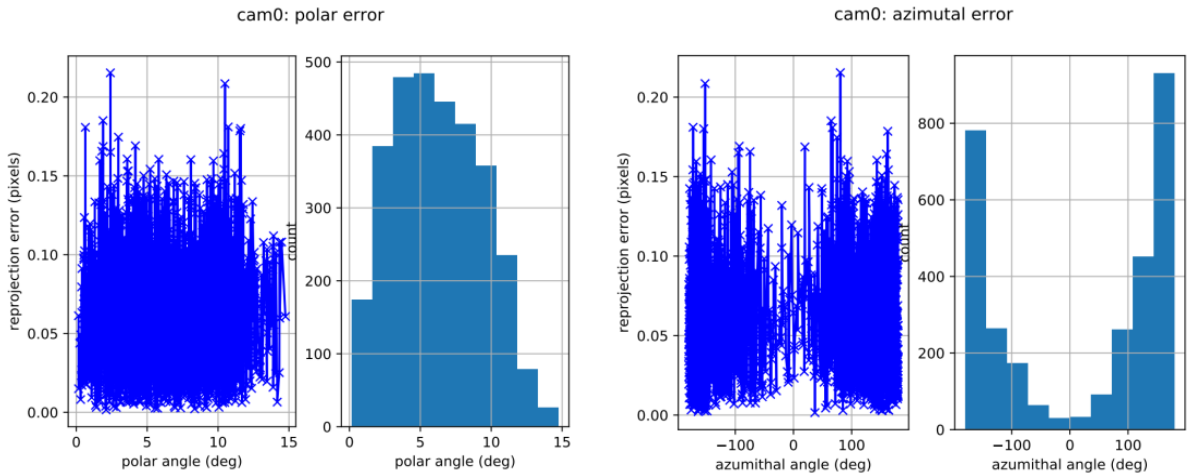

Fig. 12 Error diagram of polar angle and azimuth angle.

The reprojection error of the camera intrinsics is shown in Figure 13. It can be seen from the figure that under this calibration internal reference, the maximum reprojection error does not exceed 0.5 pixels, which is within the allowable range of 1 pixel, so this calibration is valid and it is used as the internal reference of the MYNT EYE Camera D1010-50.

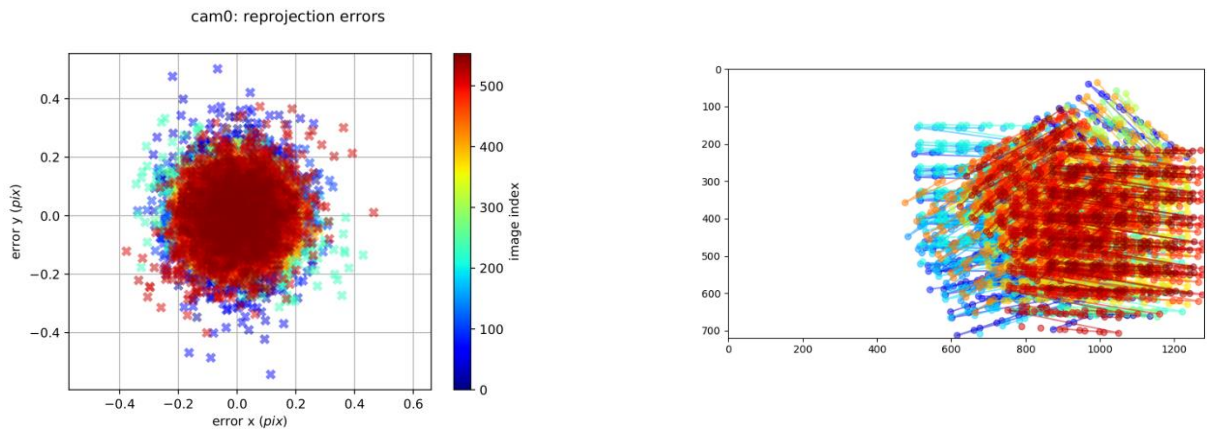

(1) Reprojected Residuals Diagram.    (2) Camera Movement Diagram.

Fig. 13 Camera reprojection error diagrams.

## 4.2 Experiment of IMU Intrinsic Parameter Calibration

In this paper, the MYNT camera was left still and recorded data packets for 120 minutes, in which the data volume of the recorded accelerometer and angular velocity meter on the xyz axis was 1358672. The calculated IMU internal parameters are shown in Table 3.

Table 3 IMU random error calibration results

| Category | Axis | Values |
|---|---|---|
| Gyro Noise (unit: rad/s) | $x$ | 1.7150288432418979e-03 |
| | $y$ | 2.2810285332151279e-03 |
| | $z$ | 1.4803156902749898e-03 |
| Accelerometer Noise (unit: m/s$^2$) | $x$ | 2.0263473443483032e-02 |
| | $y$ | 1.9452407902823307e-02 |
| | $z$ | 2.0800966656033096e-02 |
| Gyro Random Walk (unit: rad/s) | $x$ | 3.0555539333903460e-05 |
| | $y$ | 4.6905078690435216e-05 |
| | $z$ | 9.4998484143845586e-06 |
| Accelerometer Random Walk (unit: m/s$^2$) | $x$ | 3.3002849804958574e-04 |
| | $y$ | 3.5835234450158800e-04 |
| | $z$ | 3.3851545364446103e-04 |

The average value of gyroscope noise, random walk, accelerometer noise, and random walk can be calculated from the above table, as shown in Table 4.

Table 4 The average value of IMU calibration results

| Category | Parameters | Values |
|---|---|---|
| Gyro (unit: rad/s) | Gyro Noise | 1.8254576889106717e-03 |
| | Gyro Random Walk | 2.8986822146241081e-05 |
| Accelerometer (unit: m/s$^2$) | Accelerometer Noise | 2.0172282667446476e-02 |
| | Accelerometer Random Walk | 3.4229876539854489e-04 |

The Allan variance method was proposed by David Allan of the National Bureau of Standards in the 1960s, which is an analysis method based on time domain [17]. Its main feature is the ease of detailed characterization and identification of various error sources and their contributions to the overall statistical properties of the noise. The Allan variances of the gyroscope and accelerometer when the IMU is calibrated in this paper are shown in Figures 14 and 15, respectively.

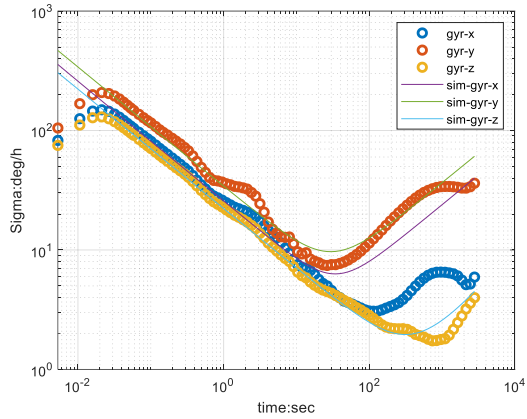
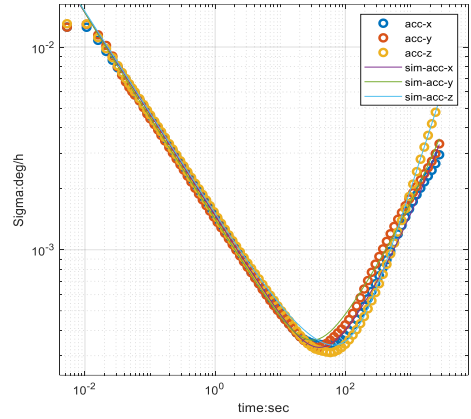

Fig. 14. Diagram of Allan Variance for gyroscope.   Fig. 15. Diagram of Allan Variance for accelerator.

For IMU deterministic error calibration, the calibration process used in this paper mainly includes: (1) put the IMU on the table and let it stand for 50 seconds; (2) draw the IMU in the air and rotate it; (3) put the IMU on the Put it on the desktop for 1 second, and repeat the above steps about 50 times to complete the calibration. The acceleration and time interval on the *xyz* three axes during the calibration in this paper are shown in Figure 16.

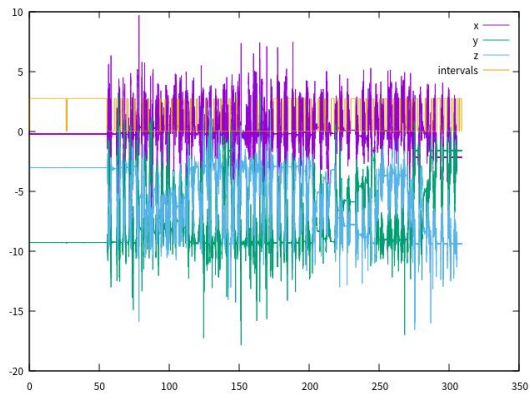

Fig. 16. IMU *xyz* three-axis acceleration and time interval diagram.

The calculated parameters related to the IMU deterministic error are shown in Table 5.

Table 5 IMU deterministic error calibration results

| Category | Parameters | Values |
| --- | --- | --- |
|  | $x$-axix | 14.1412 |
| Accelerometer Bias | $y$-axix | 7.51009 |
|  | $z$-axix | 4.37088 |
| Gyroscope Bias | $x$-axix | 14.1412 |
| | $y$-axix | 7.51009 |
| | $z$-axix | 4.37088 |
| Accelerometer Scale Matrix | $\boldsymbol{S}_{acc} = \begin{bmatrix} 0.636032 & 0 & 0 \\ 0 & 0.592957 & 0 \\ 0 & 0 & 0\ 0.666332 \end{bmatrix}$ | |
| Accelerometer Offset Coefficient Matrix | $\boldsymbol{M}_{acc} = \begin{bmatrix} 1 & -0.744004 & -0.3843 \\ 0 & 1 & -0.297786 \\ 0 & 0 & 1 \end{bmatrix}$ | |
| Gyroscope Scale Matrix | $\boldsymbol{S}_{acc} = \begin{bmatrix} 0.636032 & 0 & 0 \\ 0 & 0.592957 & 0 \\ 0 & 0 & 0\ 0.666332 \end{bmatrix}$ | |
| Gyroscope Offset Coefficient Matrix | $\boldsymbol{M}_{acc} = \begin{bmatrix} 1 & -0.744004 & -0.3843 \\ 0 & 1 & -0.297786 \\ 0 & 0 & 1 \end{bmatrix}$ | |

## 4.3 Experiment of Camera-IMU Extrinsic Parameter Calibration

In order to verify the online calibration method of the camera and IMU extrinsic parameters proposed in this paper, the Aprilgrid calibration board is fixed in this paper, and the data packets of the movement, rotation, and "8" of the calibration board are recorded by the MYNT camera through ROS. We totally collected 1858 image data for this experiment. The acceleration error and angular velocity error in the calibration process are shown in Figure 17-18.

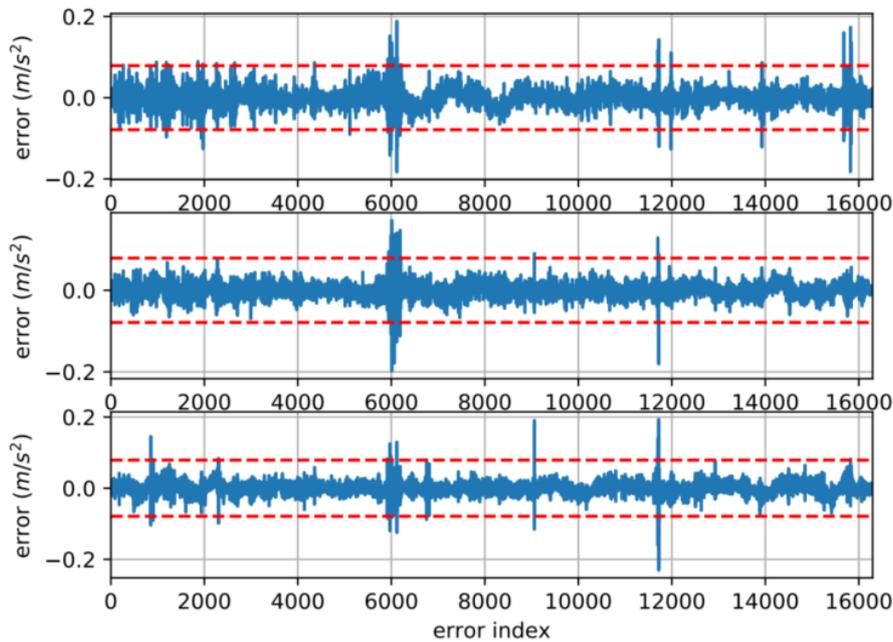

Fig. 17. Acceleration error graph.

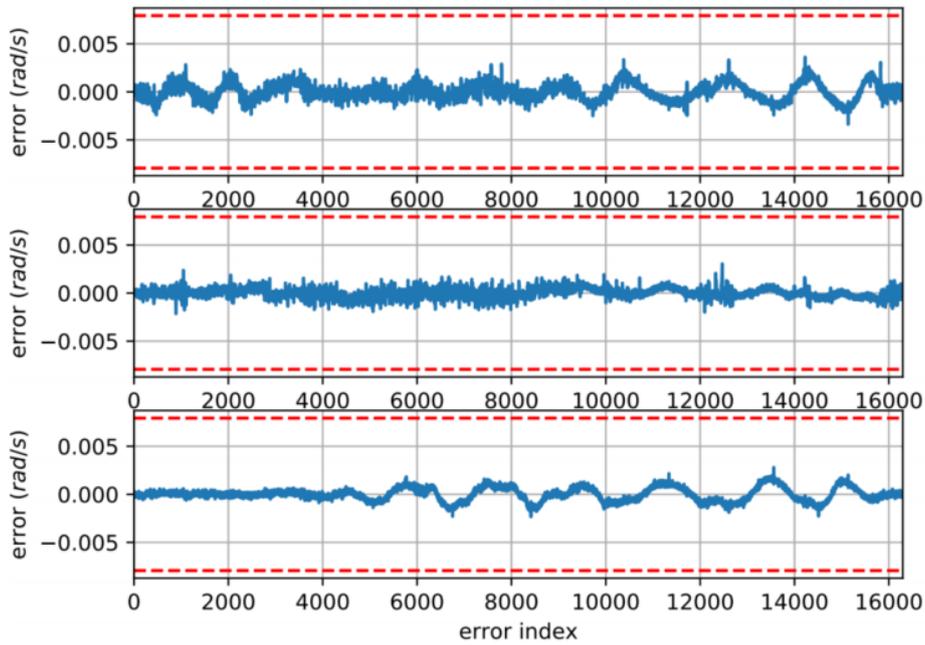

Fig. 18. Angular velocity error graph.

It can be seen from the figure that the acceleration error and angular velocity error are basically within the allowable range. The estimated accelerometer and gyroscope biases are shown in Figure 19 and Figure 20.

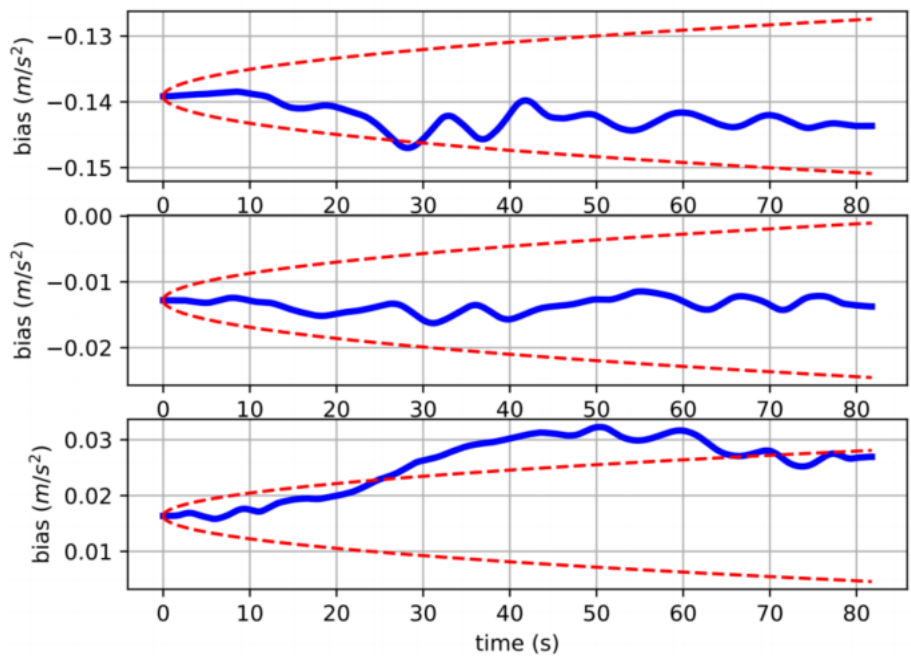

Fig. 19. Accelerometer bias estimation graph.

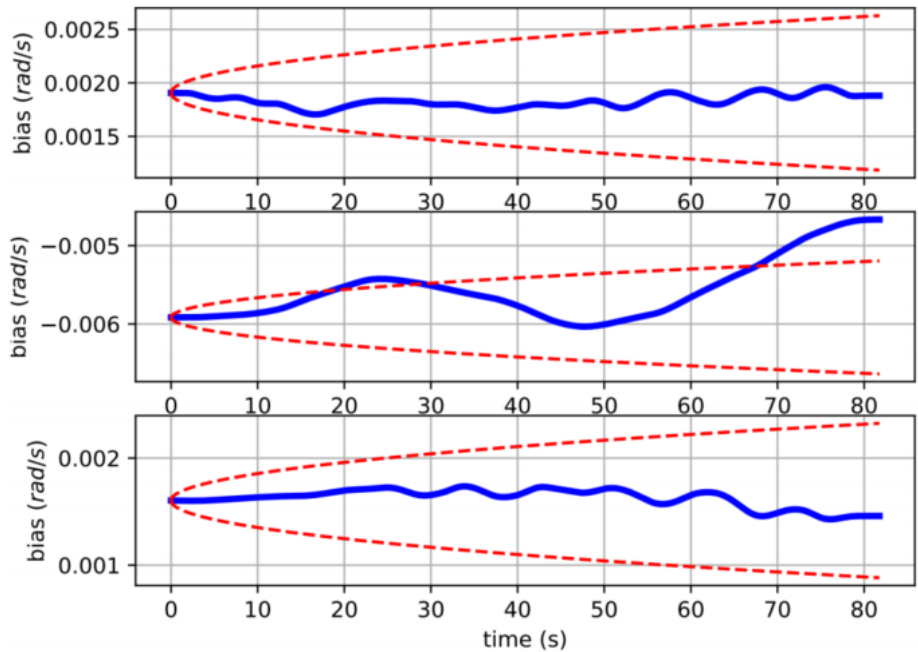

Fig. 20. Gyroscope bias estimation graph.

    It can be seen from the figure that the estimated values of accelerometer and gyroscope bias are basically within a reasonable range, and only a few times the estimated values exceed the range. This is mainly because the MYNT camera used in this article is a low-cost sensor, and its IMU has low precision . Figure 21 and Figure 22 compare the predicted and measured values of specific force and angular velocity, and Figure 23 shows the reprojection error of the camera.

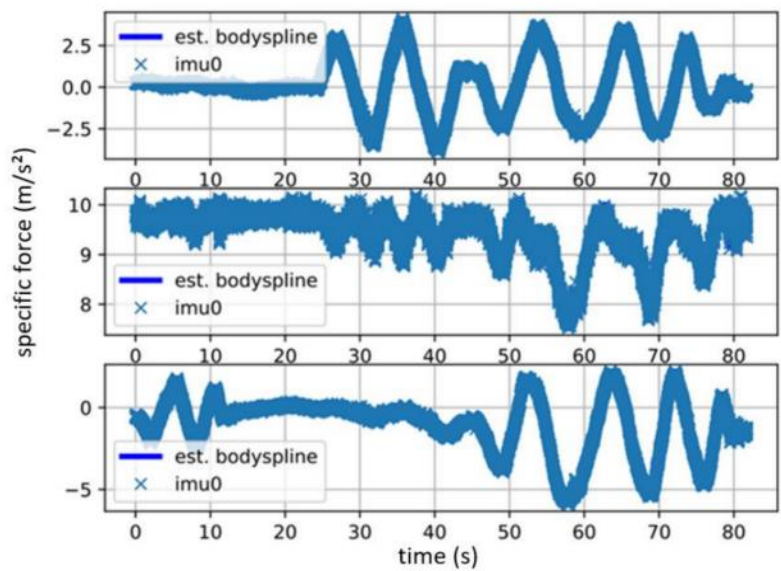

Fig. 21. Comparison graph of predicted and measured specific force.

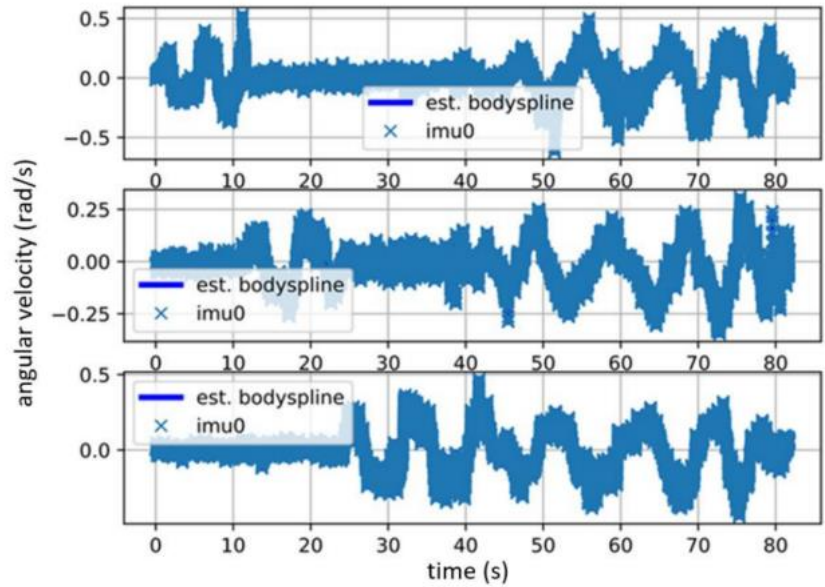

Fig. 22. Comparison graph of predicted and measured angular velocity.

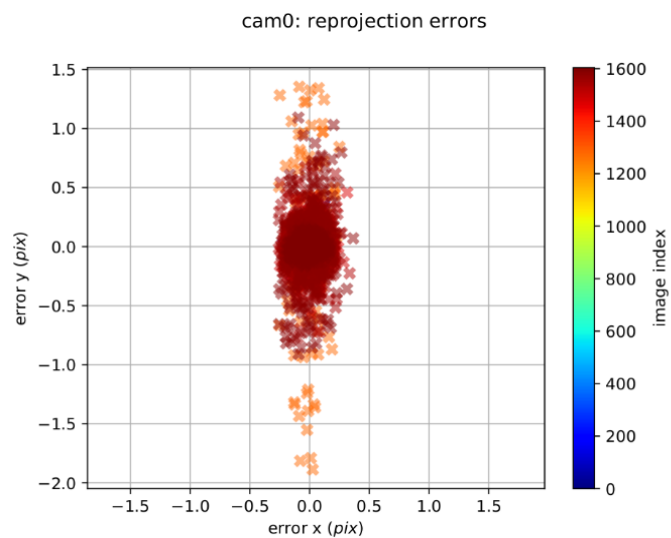

Fig. 23. Camera reprojection error diagram.

It can be seen from the figure that the predicted and measured values of specific force and angular velocity are basically consistent and the camera reprojection error is basically within 1 pixel. Therefore, the calibrated extrinsic parameters is considered to be accurate and the proposal methods in this paper are reliable. The rotation matrix $\boldsymbol{R}_{bc}$ and translation matrix $\boldsymbol{t}_{bc}$ from the calibrated camera coordinate system to the IMU coordinate system are as follows.

$$\boldsymbol{R}_{bc} = \begin{bmatrix} 0.99976023688788251 & 0.012861567340825 & -0.0177214227256832 \\ 0.0137597595321047 & -0.998576606647093 & -0.0515308613822358 \\ -0.017033430526573 & 0.0517623486978128 & -0.99851415688601164 \end{bmatrix} \quad (47)$$

$$\boldsymbol{t}_{bc} = [0.05715507571041868 \quad 0.254279370393975 \quad 0.01424385318401530]^T \tag{48}$$

## 5. Conclusion

In this paper, we present a fast and robust camera-IMU online calibration method based space coordinate transformation constraints and SVD (singular Value Decomposition) tricks. First, constraint equations are constructed based on equality of rotation and transformation matrices between camera frames and IMU coordinates at different moments. Secondly, the external parameters of the camera-IMU are solved using quaternion transformation and SVD techniques. The experimental results demonstrated the performance of our proposed Camera-IMU online calibration method. In the future, we will consider the characteristic of VSLAM system to further model the measurement and noise of camera and IMU sensors.

## References

1. Zhu B, Tao X, Zhao J, et al. An integrated GNSS/UWB/DR/VMM positioning strategy for intelligent vehicles[J]. IEEE Transactions on Vehicular Technology, 2020, 69(10): 10842-10853.
2. Zhu B, Zhang P, Zhao J, et al. Hazardous scenario enhanced generation for automated vehicle testing based on optimization searching method[J]. IEEE Transactions on Intelligent Transportation Systems, 2021, 23(7): 7321-7331.
3. Li X, Tao X, Zhu B, et al. Research on a simulation method of the millimeter wave radar virtual test environment for intelligent driving[J]. Sensors, 2020, 20(7): 1929.
4. Tao X, Zhu B, Xuan S, et al. A multi-sensor fusion positioning strategy for intelligent vehicles using global pose graph optimization[J]. IEEE Transactions on Vehicular Technology, 2021, 71(3): 2614-2627.
5. Bing Zhu, Xiaowen Tao, Jian Zhao, et al. Two-stage UWB positioning algorithm of intelligent vehicle[J]. Journal of Traffic Engineering, 2021, 21(2): 256-266.
6. Zhang Z. A flexible new technique for camera calibration[J]. IEEE Transactions on pattern analysis and machine intelligence, 2000, 22(11): 1330-1334.
7. Heikkila J, Silvén O. A four-step camera calibration procedure with implicit image correction[C]//Proceedings of IEEE computer society conference on computer vision and pattern recognition. IEEE, 1997: 1106-1112.
8. Salvi J, Armangué X, Batlle J. A comparative review of camera calibrating methods with accuracy evaluation[J]. Pattern recognition, 2002, 35(7): 1617-1635.
9. Zhang Z, Scaramuzza D. A tutorial on quantitative trajectory evaluation for visual (-inertial) odometry[C]//2018 IEEE/RSJ International Conference on Intelligent Robots and Systems (IROS). IEEE, 2018: 7244-7251.
10. Brink K, Soloviev A. Filter-based calibration for an IMU and multi-camera system[C]//Proceedings of the 2012 IEEE/ION Position, Location and Navigation Symposium. IEEE, 2012: 730-739.
11. Sünderhauf N. An Outlook on Robust Optimization for Sensor Fusion and Calibration[M]//Switchable Constraints for Robust Simultaneous Localization and Mapping and Satellite-Based Localization. Cham: Springer International Publishing, 2023: 167-170.
12. Zhu B, Zhang P, Zhao J, et al. Hazardous scenario enhanced generation for automated vehicle testing based on optimization searching method[J]. IEEE Transactions on Intelligent Transportation Systems, 2021, 23(7): 7321-7331.
13. Zhao J, Li Y, Zhu B, et al. Method and applications of LiDAR modeling for virtual testing of intelligent vehicles[J]. IEEE Transactions on Intelligent Transportation Systems, 2020, 22(5): 2990-3000.
14. Zhu B, Han J, Zhao J, et al. Combined hierarchical learning framework for personalized automatic lane-changing[J]. IEEE Transactions on Intelligent Transportation Systems, 2020, 22(10): 6275-6285.
15. Han J, Zhao J, Zhu B, et al. Adaptive steering torque coupling framework considering conflict resolution for human-machine shared driving[J]. IEEE Transactions on Intelligent Transportation Systems, 2021, 23(8): 10983-10995.
16. Zhang P, Zhu B, Zhao J, et al. Performance evaluation method for automated driving system in logical scenario[J]. Automotive Innovation, 2022, 5(3): 299-310.
17. Jiang Y, Zhu B, Yang S, et al. Vehicle Trajectory Prediction Considering Driver Uncertainty and Vehicle Dynamics Based on Dynamic Bayesian Network[J]. IEEE Transactions on Systems, Man, and Cybernetics: Systems, 2022.